\documentclass{amsart}
\usepackage[utf8]{inputenc}
\usepackage{amsmath}
\usepackage{mathtools}
\usepackage{amsfonts}
\usepackage{amssymb,amsthm}
\usepackage{amsaddr}

\title[MaxTSP in one dimension]{Exact cost of the  Euclidean Maximum Traveling Salesman Problem in one dimension}

\author{Sergio Caracciolo}
\address{INFN, Sezione di Milano, via Celoria 16, I-20133 Milan, Italy} 
\email{sergio.caracciolo@mi.infn.it}
\author{Carlo Vanoni}
\address{Department of Physics, Princeton University, Princeton, NJ 08544, USA}
\email{cvanoni@princeton.edu}
\date{\today}

\numberwithin{equation}{section}

\usepackage{booktabs}
\usepackage{graphicx}
\usepackage{ifthen,xspace,subfig,wrapfig,rotating,listings}
\usepackage{tikz,pgfplots,pgfplotstable,tkz-berge,tkz-graph}
\usepackage{bbm,bm,mathrsfs,dsfont}

\usepackage{enumitem}

\usepackage[numbers,sort&compress]{natbib}

\usepackage{hyperref}

\newcommand{\be}{\begin{equation}}
\newcommand{\ee}{\end{equation}}
\newtheorem{thm}{Theorem}[section]

\newtheorem{lem}[thm]{Lemma}

\newcommand{\m}{\rm min}
\newcommand{\M}{\rm max}

\begin{document}

\maketitle

\begin{abstract}
    We study the random Euclidean Maximum Traveling Salesman Problem: finding a Hamiltonian cycle of maximal total cost through any set of $N$ cities on a line, where each edge cost is a function of its Euclidean length. We characterize the maximizing cycles for convex increasing, concave increasing, and convex decreasing cost functions. For points sampled independently and uniformly from the unit interval, we derive explicit expressions for the average maximal cost for arbitrary $N$ and asymptotic expansions with power-law edge costs.
\end{abstract}

\section{Introduction}

The average properties of random combinatorial optimization problems can be mapped onto the statistical mechanics of disordered systems at zero temperature~\citep{Kirkpatrick1981,Kirkpatrick1983,KS,Sourlas1986,FA,BSS}. This equivalence has been extensively leveraged for random problems defined on graphs where edge weights are independent, identically distributed random variables~\citep{Vannimenus1984,Orland1985,Mezard1985,Mezard1986,Mezard1986a,Mezard1987,mezard1987spin,Krauth1989,mezard2009information}, a regime where spatial geometry exerts no influence.

Conversely, in Euclidean optimization problems, the graph is embedded in a compact subspace $S\subset\mathbb{R}^d$. Here, while the positions of the $N$ vertices $x_1,\dots,x_N \in S$ remain independent random variables, the weights assigned to the edges connecting vertices $i$ and $j$ (with $i,j \in [N] =\{1,\dots,N\}$) are given by:
\be w_{ij} =\phi (|x_i-x_j|). \ee
Because these weights depend on the Euclidean distance between nodes, they are
correlated through the geometry~\cite{Ajtai,Talagrand_92,Talagrand1992,Talagrand94a,Talagrand94b,Yukich}. These correlations break the independence assumption, preventing the use of standard replica or cavity methods to derive the typical properties of the system. Only in the limit of very high dimensions $d$ can these geometric correlations be treated perturbatively around the random-link solution~\cite{Mezard1988}. This transition closely mirrors the shift from a mean-field solution to a finite-dimensional model in classical statistical mechanics.
At the other extreme, the one-dimensional limit $d=1$ is often simple enough to yield exact analytical solutions. Indeed, McCann dedicated his work~\cite{McCannRobert1999} on the $d=1$ assignment problem to Elliott Lieb, whose pioneering contributions to 1D statistical mechanics~\cite{schultz1961two,lieb1963exact,lieb1968absence} have had a profound impact on the field.
For several random Euclidean optimization problems on complete 1D graphs, exact results are available—provided the cost function is convex with respect to the edge lengths. These include the matching problem (or 1-factor / dimer covering)~\cite{McCannRobert1999,Caracciolo:159,Caracciolo:160,Caracciolo:169,Sicuro,Caracciolo:172,Caracciolo:177}, the 2-factor problem (or 2-matching / loop covering)~\cite{Caracciolo:173,Caracciolo:177}, and the traveling salesman problem (a Hamiltonian cycle visiting every vertex)~\cite{Caracciolo:171,Caracciolo:177}. Interestingly, the optimal total costs in all these problems are non-self-averaging and obey simple, explicit relationships. Furthermore, insights gained from the \(d=1\) solutions have paved the way for extensions to higher dimensions, at least under specific choices of the cost function~\cite{Caracciolo:158,Caracciolo:162,Caracciolo:163,Ambrosio2016,Ambrosio2019,Caracciolo:174,Caracciolo:182,Ambrosio24,Matteo}. 
 
In this paper, we investigate the Maximum Euclidean Traveling Salesman Problem (MaxTSP) in one dimension. The MaxTSP~\citep{Supnick,Supnick_64,Quintas_65,Kalmanson} is also colloquially known as the “taxi ripoff problem," an allusion to the real-world fraud where a taxi driver deliberately takes an unnecessarily long route to inflate the fare (“long-hauling"). See also the more recent works~\cite{Deneko,DEINEKO,OKAMOTO2004349,Barvinok_2007}.
We call our MaxTSP {\em Euclidean} because we consider a cost function for the edge connecting the points at positions $x_i$ and $x_j$, with $i,j\in [N] =\{1,\dots,N\}$, given by
\be
w_{ij} = \phi(|x_i-x_j|)
\ee
where $|x_i-x_j|$ is their Euclidean distance and $\phi: \mathbb{R}^+\to \mathbb{R}^+$ is a function to be specified.

This work can be seen as the counterpart of paper~\citep{Caracciolo:178}, where the  Euclidean TSP, with minimal cost, on the complete graph (MinTSP),  has been studied in $d=1$, with a cost function 
\be
\phi(x) = x^p
\ee 
for various values of $p\in \mathbb{R}$. 
There are three distinct regions of interest:
\begin{enumerate}[label=(\roman*)]
\item[i)] $p>1$ where the cost function $\phi$ is convex and increasing;
\item[ii)] $0<p<1$ where the cost function $\phi$ is concave and increasing;
\item[iii)] $p<0$ where the cost function $\phi$ is convex and decreasing.
\end{enumerate}
In the cases $p=0$ and $p=1$, the solutions are degenerate. Of course, for $p=0$ all the configurations have the same weight, that is $N$, the number of steps in any Hamiltonian cycle. In the Euclidean assignment problem in $d=1$, the interested reader can find the exact counting of the degenerate solutions for $p=1$ in~\cite{Caracciolo:181}.

Of course, if the cost function is not restricted to be positive, the search for the MaxTSP is the same as the search for the MinTSP with cost function $-\phi$, a transformation that maps a convex function to a concave one and an increasing function to a decreasing one. We immediately remark that the properties of the function $\phi$ in the regions $0<p<1$ and $p<0$ are exchanged by this involution. So the solutions of the MinTSP and MaxTSP will also be exchanged.

After a more formal introduction to the problem in Section~\ref{problem}, we review, in Section~\ref{Lemmas}, comparison Lemmas for the matchings of a quadrupole of points which are essential in the search for the optimal Hamiltonian cycles, in a form which is suitable for our aims. The cases in which the cost function is concave and increasing, respectively convex and decreasing, for the MaxTSP are related to the MinTSP in the exchanged cases. We, therefore, have adapted the results in~\cite{Caracciolo:178} for our problem in Section~\ref{solpl1}. The case in which the cost function is convex and increasing is examined in detail in Section~\ref{solpg1}. For $p>1$, the Hamiltonian cycle that maximizes the cost is unique for each number of points. Their form is explicitly presented for the cases in which $N$ is even or odd.
In Section~\ref{AOC} we use the previous results to give the expression of the average cost of the MaxTSP for the random version of the problem. Explicit values are derived in the case in which the $N$ points are chosen uniformly, and the cost functions are of the form $\phi(x)=x^p$. Some conclusions are given in Section~\ref{end}. In addition, we give in Appendix~\ref{AppA} the essential technical ingredient to see that the average cost is a rational function of $N$ when $p$ is a natural number. In Appendix~\ref{AppB} we show how to derive the asymptotic behaviors of our expression for a large number of points.

\section{The problem}\label{problem}

A generic (undirected) graph $\mathcal{G}$ is the pair $(\mathcal{V},\mathcal{E})$, where $\mathcal{V}$, or more explicitly $\mathcal{V}(\mathcal{G})$, is its vertex set and $\mathcal{E}$, or more explicitly $\mathcal{E(}\mathcal{G})$, is its edge set. 

A graph $\mathcal{G}$ is weighted when there is a mapping from $\mathcal{E}$ to $\mathbb{R}^+$ which assigns to each edge $e\in \mathcal{E}$ a cost $w_e \geq 0$.

A {\em cycle}  of {\em length} $k$  is a sequence of edges $e_1,\dots,e_k\in \mathcal{E}$ in which two subsequent edges $e_j$ and $e_{j+1}$ for $j=1, \dots, k-1$, together with the pair $e_k$ and $e_1$, share a common vertex in $\mathcal{V}$, which is said to be {\em visited} by the cycle. A cycle is said to be {\em simple} when each visited vertex is visited only once. A {\em Hamiltonian} cycle is a simple cycle which visits all the vertices in $\mathcal{V}$. Let $\mathcal{H}$ be the set of Hamiltonian cycles.

If $|\mathcal{V}|=N$, we can identify the vertices with their labels in $[N]=\{1,\dots,N\}$, that is, set $\mathcal{V}=[N]$. Then the edge in $\mathcal{E}$ that connects the vertices $i$ and $j$ will be denoted by $\{i,j\}$, and its weight $w_e$ with $w_{ij}$.

To each Hamiltonian cycle $h\in\mathcal{H}$ we associate a cost
\be
E(h) \coloneqq \sum_{e\in h } w_e  .
\ee

The search for the Hamiltonian cycle $\hat{h}_{\rm min}$ with minimum cost
\be
E_{\m} = \min_{h\in \mathcal{H}} E(h) = E(\hat{h}_{\m})
\ee
is called the {\em minimum Traveling Salesman Problem}, a classical problem in {\em combinatorial optimization}.

Similarly, the search for the Hamiltonian cycle $\hat{h}_{\rm max}$ with maximum cost
\be
E_{\M} = \max_{h\in \mathcal{H}} E(h) = E(\hat{h}_{\M})
\ee
is called the {\em maximum Traveling Salesman Problem} (see for example~\citep{Barvinok_2003,Barvinok_2007}).

When the graph $\mathcal{G}$ is embedded in $\mathbb{R}^d$, to each vertex $i$ is associated a point $x_i \in \mathbb{R}^d$. When there is a  function $\phi: \mathbb{R}^+\to \mathbb{R}^+$ such that
\be
w_{ij} = \phi (|x_i-x_j|) ,
\ee
that is, the weight of the edge $\{i,j\}$ is a given function of the Euclidean distance between the points $x_i$ and $x_j$,
we talk about the {\em Euclidean} version of an optimization problem.

We shall consider the {\em complete graph} $\mathcal{K}_N$ with $N$ vertices. For simplicity, we shall take its vertex set as $[N]=\{1,2,\dots,N\}$. Let each vertex  $i\in [N]$ have a position $x_i\in I$, where $I$ is an interval of the real line. Assume also that all the positions are different from each other. To be definite, we shall take $I=[0,1]$. Let us choose a re-labelling of the vertices so that the positions are ordered, that is
$$ x_1< x_2<\dots <x_N .$$

A {\em cyclic} permutation is a permutation consisting of a single cycle. In cycle notation, cyclic permutations are denoted by the list of their elements enclosed in a unique couple of parentheses, in the order in which they are permuted. Under this definition, each permutation $\sigma\in\mathcal{S}_N$ defines a cyclic permutation, but cyclic shifts of this list define the same cycle so we may fix the first entry. This means that there are $(N-1)!$ cyclic permutations in $\mathcal{S}_N$. Reversing the list, which corresponds to the inverse permutation $\sigma^{-1}$, defines the same {\em undirected} cycle. Thus, for $N\geq3$, the number of Hamiltonian cycles on $\mathcal{K}_N$ is $N!/(2N)$.

\section{The main Lemmas}\label{Lemmas}

The main ingredient we need is a comparison theorem on a quadrupole of vertices, similar to those already used, for example, in~\citep{McCannRobert1999,Caracciolo:159,Caracciolo:160,Caracciolo:169,Caracciolo:171,Caracciolo:178}. 
Given four points $a,b,c,d$ in the interval $[0,1]$ with $a\leq b\leq c\leq d$, there are three possible perfect matchings:
\begin{enumerate}
\item $M_1=\{ \{a,b\}, \{c,d\}\}$ which we call {\em parallel};
\item $M_2 = \{ \{a,c\}, \{b,d\}\}$ which we call {\em crossing}; indeed, when the two arcs are drawn on the same side, they do cross;
\item $M_3 = \{ \{a,d\}, \{b,c\}\}$ which we call {\em nested}
\end{enumerate}

\newcommand{\arcbelow}[2]{%
  \draw (#1,0) arc[start angle=180, end angle=360, radius={(#2-#1)/2}];}

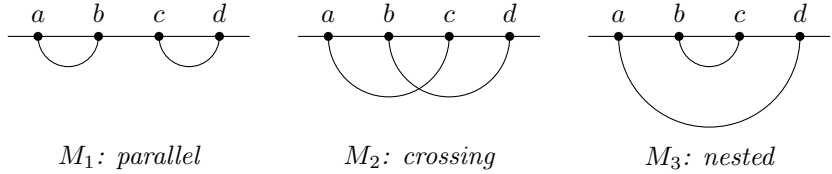
\begin{figure}[ht]
  \centering
  \begin{tikzpicture}[
    scale=0.8,
    dot/.style={circle, fill, inner sep=1.2pt},
    lbl/.style={font=\itshape}
  ]
  \begin{scope}[xshift=0cm]
    \draw (-0.5,0) -- (3.5,0);
    \arcbelow{0}{1}
    \arcbelow{2}{3}
    \foreach \x/\l in {0/a,1/b,2/c,3/d} \node[dot, label=above:{$\l$}] at (\x,0) {};
    \node[lbl] at (1.5,-2) {$M_1$: parallel};
  \end{scope}

  \begin{scope}[xshift=4.8cm]
    \draw (-0.5,0) -- (3.5,0);
    \arcbelow{0}{2}
    \arcbelow{1}{3}
    \foreach \x/\l in {0/a,1/b,2/c,3/d} \node[dot, label=above:{$\l$}] at (\x,0) {};
    \node[lbl] at (1.5,-2) {$M_2$: crossing};
  \end{scope}
  
  \begin{scope}[xshift=9.6cm]
    \draw (-0.5,0) -- (3.5,0);
    \arcbelow{0}{3}
    \arcbelow{1}{2}
    \foreach \x/\l in {0/a,1/b,2/c,3/d} \node[dot, label=above:{$\l$}] at (\x,0) {};
    \node[lbl] at (1.5,-2) {$M_3$: nested};
  \end{scope}

  \end{tikzpicture}
  \caption{Parallel, crossing, and nested matching configurations.}
  \label{fig:arcs}
\end{figure}
Given the cost function $\phi:[0,1] \to \mathbb{R}^+$, their costs will be
\begin{align}
E[M_1] = & \phi(b-a) + \phi(d-c)\\
E[M_2] = & \phi(c-a) + \phi(d-b)\\
E[M_3] = & \phi(d-a) + \phi(c-b)
\end{align}

\begin{lem}
If the cost function $\phi:[0,1] \to \mathbb{R}$ is convex, then the cost of the nested matching is higher than the cost of the crossing one:
\begin{equation} 
E[M_3] \geq E[M_2].
\end{equation}
The inequality is strict if \( \phi \) is strictly convex and the points are strictly ordered. 
If, instead, $\phi$ is concave, the inequality is reversed.
\end{lem}

\begin{proof}
The convexity of $\phi$ implies that the map
$$ g(\delta) = \phi(x+\delta) - \phi(\delta) $$
is nondecreasing for $\delta\geq 0$ and fixed $x\geq 0$. Indeed, take $0\leq \delta_1 \leq \delta_2$; as $\phi$ is convex, its rate of change in the interval $[\delta_1, \delta_1+x]$ must be less than its rate of change in the interval $[\delta_2, \delta_2+x]$
$$\frac{\phi(x+\delta_1) - \phi(\delta_1)}{x} \leq \frac{\phi(x+\delta_2) - \phi(\delta_2)}{x}$$
and since $x\geq 0$
$$\phi(x+\delta_1) - \phi(\delta_1) \leq \phi(x+\delta_2) - \phi(\delta_2). $$
Let $u= b-a$, $v= c-b$, $w= d-c$, and set $\delta_1= v \leq \delta_2 = u+v, x = w$. Then
$$\phi(u+v+w) - \phi(u+v) \geq \phi(w+v) - \phi(v). $$
That is 
$$ \phi(d-a) - \phi(c-a) \geq \phi(d-b) - \phi(c-b)$$
which is to say
$$ \phi(d-a) + \phi(c-b) \geq \phi(c-a) + \phi(d-b). $$
\end{proof}

\begin{lem}
If the cost function $\phi:[0,1] \to \mathbb{R}$ is increasing, then the cost of the crossing matching is higher than the cost of the parallel one:
\begin{equation} 
E[M_2] \geq E[M_1].
\end{equation}
The inequality is strict if \( \phi \) is strictly increasing and the points are strictly ordered. 
If, instead, $\phi$ is decreasing, the inequality is reversed.
\end{lem}

\begin{proof}
Let $u=b-a$, $v=c-b$, $w=d-c$. We have $u, v, w\geq 0$.

\begin{align}
\phi(c-a) + \phi(d-b) & = \phi(u+v) + \phi(w+v) \\
                                & \geq \phi(u) + \phi(w) \\
                                & = \phi(b-a) + \phi(d-c) .
\end{align}
\end{proof}

According to these Lemmas, we have that 
\begin{enumerate}[label=(\roman*)]
\item[i)] when the cost function $\phi$ is convex and increasing, which, in the case $\phi(x)=x^p$, corresponds to the region $p>1$, the three matchings are completely ordered by their costs:
\be
E[M_3] \geq E[M_2] \geq E[M_1] ;
\ee
\item[ii)] when the cost function $\phi$ is concave and increasing, which corresponds to the region $0<p<1$, the crossing matching has the highest cost
\be
E[M_2] \geq E[M_3] \ \qquad E[M_2] \geq E[M_1]
\ee
\item[iii)] when the cost function $\phi$ is convex and decreasing, which corresponds to the region $p<0$, the crossing matching has the lowest cost
\be
E[M_3] \geq E[M_2] ; \qquad E[M_1] \geq E[M_2]. 
\ee
\end{enumerate}

\section{The explicit solutions for $p<1$}\label{solpl1}

Let's consider the problem studied in ~\citep{Caracciolo:178}. In that paper, the optimal solutions for the Monopartite MinTSP in 1D are obtained. It has been proven that when $0<p<1$, that is, when the crossing matching has the higher cost, the MinTSP, for a generic number of points $N$, is given by the cyclic permutation $(1,2,\ldots,N)$, that is
\be
\hat{h}_{\m}(i) = i 
\ee
which has no crossing at all. Therefore, the identity cycle, which minimizes the cost for \(0<p<1\), maximizes it for \(p<0\).
\begin{figure}[h]
		\begin{center}
			\begin{tikzpicture}[scale=0.6]
			\node[draw,circle,inner sep=1.5pt,fill=black,text=white,label=left:{\footnotesize $r_1$}] (r1) at (1,3) {};
			\node[draw,circle,inner sep=1.5pt,fill=black,text=white,label=left:{\footnotesize $r_2$}] (r2) at (3,3) {};
			\node[draw,circle,inner sep=1.5pt,fill=black,text=white,label=left:{\footnotesize $r_3$}] (r3) at (5,3) {};
			\node[draw,circle,inner sep=1.5pt,fill=black,text=white,label=left:{\footnotesize $r_4$}] (r4) at (7,3) {};
			\node[draw,circle,inner sep=1.5pt,fill=black,text=white,label=right:{\footnotesize $r_5$}] (r5) at (9,3) {};
			\node[draw,circle,inner sep=1.5pt,fill=black,text=white,label=right:{\footnotesize $r_6$}] (r6) at (11,3) {};
			\node[draw,circle,inner sep=1.5pt,fill=black,text=white,label=right:{\footnotesize $r_7$}] (r7) at (13,3) {};
			\node[draw,circle,inner sep=1.5pt,fill=black,text=white,label=right:{\footnotesize $r_8$}] (r8) at (15,3) {};
			\draw[line width=1pt,gray]  (r1) to[in=+90, out=+90] (r2);
			\draw[line width=1pt,gray]  (r3) to[in=+90, out=+90] (r2);
			\draw[line width=1pt,gray]  (r4) to[in=+90, out=+90] (r3);
			\draw[line width=1pt,gray]  (r4) to[in=+90, out=+90] (r5);
			\draw[line width=1pt,gray]  (r5) to[in=+90, out=+90] (r6);
			\draw[line width=1pt,gray]  (r6) to[in=+90, out=+90] (r7);
			\draw[line width=1pt,gray]  (r8) to[in=+90, out=+90] (r7);
			\draw[line width=1pt,gray]  (r1) to[in=+90, out=+90] (r8);
			\end{tikzpicture}
		\end{center}
		\caption{MinTSP solution for $0<p<1$ and MaxTSP solution for $p<0$ when $N=8$}
\end{figure}
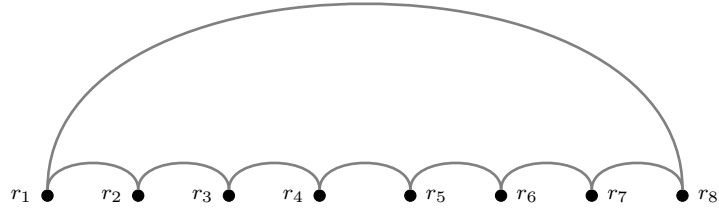

When $p<0$, that is, when the crossing matching has the lowest cost, we have to distinguish between the cases in which $N$ is even or odd.
When $N=2k-1$ is odd
\be
\hat{h}_{\m}(i)  =\begin{cases}	1 & \mbox{for } i=1 \\ k + 2 - \frac{i}{2}  &\mbox{for even } i>1 \\ 2k -\frac{i+1}{2}  &\mbox{for odd } i>1.\\ \end{cases}  
\ee
\begin{figure}[h]
		\begin{center}
			\begin{tikzpicture}[scale=0.6]
			\node[draw,circle,inner sep=1.5pt,fill=black,text=white,label=left:{\footnotesize $x_1$}] (r1) at (1,3) {};
			\node[draw,circle,inner sep=1.5pt,fill=black,text=white,label=left:{\footnotesize $x_2$}] (r2) at (3,3) {};
			\node[draw,circle,inner sep=1.5pt,fill=black,text=white,label=right:{\footnotesize $x_3$}] (r3) at (5,3) {};
			\node[draw,circle,inner sep=1.5pt,fill=black,text=white,label=right:{\footnotesize $x_4$}] (r4) at (7,3) {};
			\node[draw,circle,inner sep=1.5pt,fill=black,text=white,label=right:{\footnotesize $x_5$}] (r5) at (9,3) {};
			\node[draw,circle,inner sep=1.5pt,fill=black,text=white,label=right:{\footnotesize $x_6$}] (r6) at (11,3) {};
			\node[draw,circle,inner sep=1.5pt,fill=black,text=white,label=right:{\footnotesize $x_7$}] (r7) at (13,3) {};
			
			\draw[line width=1pt,gray]  (r1) to[in=+90, out=+90] (r4);
			\draw[line width=1pt,gray]  (r4) to[in=+90, out=+90] (r7);
			\draw[line width=1pt,gray]  (r7) to[in=+90, out=+90] (r3);
			\draw[line width=1pt,gray]  (r3) to[in=+90, out=+90] (r6);
			\draw[line width=1pt,gray]  (r6) to[in=+90, out=+90] (r2);
			\draw[line width=1pt,gray]  (r2) to[in=+90, out=+90] (r5);
			\draw[line width=1pt,gray]  (r5) to[in=+90, out=+90] (r1);	
			\end{tikzpicture}
			\caption{MinTSP for $p<0$ and MaxTSP solution for $0<p<1$ when $N=7$, which corresponds to the permutation cycle $(1,4,7,3,6,2,5)$.}
		\end{center}
\end{figure}
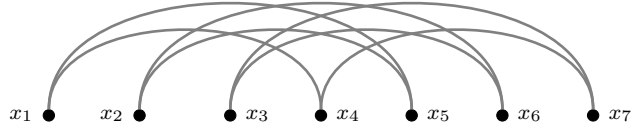
When $N=2k$ is even, there is not a unique optimal Hamiltonian cycle, but many. Each of them can be optimal under suitable choices of the positions of the points. 
Given
\begin{equation}
\bar{h}_{\m}(i) = \begin{cases}
1 & \mbox{for } i=1 \\ 
k +i-1 &\mbox{for even } i\leq k +1 \\ 
i-1  &\mbox{for odd } i\leq k+1 \\  
3 k - i +2& \mbox{for even } i > k\\ 
2k-i+2 & \mbox{for odd } i> k +1 . 
\end{cases} 
\end{equation}
Given the shift-permutation $\tau\in\mathcal{S}_N$,  defined by $\tau(i) =i+1$ for  $i\in [N-1]$ and $\tau(N)=1$, all the possible optimal Hamiltonian cycles are obtained from the set of permutations $\sigma_m(i)$,  for $m\in[N]$,  defined as $\sigma_m(i) = \tau^m (\bar{h}_{\m}(i) )$, modulo the cycles in reverse order. At the end, there are exactly $k$ distinct possible optimal Hamiltonian cycles.

For example,  when $N=4$, we get
\begin{align*}
\sigma_1(\bar{h}_{\m}) & = \bar{h}_{\m} = (1,3,2,4) =(1,4,2,3),\\
\sigma_2(\bar{h}_{\m}) & = (2,4,3,1) = (1,2,4,3),\\
\sigma_3(\bar{h}_{\m}) & = (3,1,4,2) = (1,4,2,3),\\
\sigma_4(\bar{h}_{\m}) & = (4,2,1,3) = (1,3,4,2)=(1,2,4,3),
\end{align*}
where $(1,3,2,4)$ is obtained by reversing the order of $(1,4,2,3)$, and, similarly $(1,3,4,2)$ is obtained from $(1,2,4,3)$. 

The crossing cycle for odd \(N\), and the family of crossing-cycle candidates for even \(N\), that give the MinTSP with \(p<0\), provide the MaxTSP solution and candidate family, respectively, for \(0<p<1\).
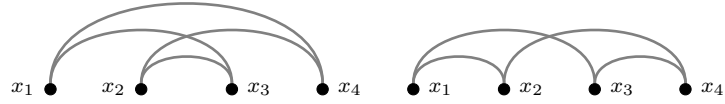
\begin{figure}[h]
	\begin{center}
	\begin{tikzpicture}[scale=0.6]
	\node[draw,circle,inner sep=1.5pt,fill=black,text=white,label=left:{\footnotesize $x_1$}] (r1) at (1,3) {};
	\node[draw,circle,inner sep=1.5pt,fill=black,text=white,label=left:{\footnotesize $x_2$}] (r2) at (3,3) {};
	\node[draw,circle,inner sep=1.5pt,fill=black,text=white,label=right:{\footnotesize $x_3$}] (r3) at (5,3) {};
	\node[draw,circle,inner sep=1.5pt,fill=black,text=white,label=right:{\footnotesize $x_4$}] (r4) at (7,3) {};
	\node[draw,circle,inner sep=1.5pt,fill=black,text=white,label=right:{\footnotesize $x_1$}] (r5) at (9,3) {};
	\node[draw,circle,inner sep=1.5pt,fill=black,text=white,label=right:{\footnotesize $x_2$}] (r6) at (11,3) {};
	\node[draw,circle,inner sep=1.5pt,fill=black,text=white,label=right:{\footnotesize $x_3$}] (r7) at (13,3) {};
	\node[draw,circle,inner sep=1.5pt,fill=black,text=white,label=right:{\footnotesize $x_4$}] (r8) at (15,3) {};
	\draw[line width=1pt,gray]  (r1) to[in=+90, out=+90] (r3);
	\draw[line width=1pt,gray]  (r3) to[in=+90, out=+90] (r2);
	\draw[line width=1pt,gray]  (r2) to[in=+90, out=+90] (r4);
	\draw[line width=1pt,gray]  (r4) to[in=+90, out=+90] (r1);
	\draw[line width=1pt,gray]  (r5) to[in=+90, out=+90] (r7);
	\draw[line width=1pt,gray]  (r7) to[in=+90, out=+90] (r8);
	\draw[line width=1pt,gray]  (r8) to[in=+90, out=+90] (r6);	
	\draw[line width=1pt,gray]  (r6) to[in=+90, out=+90] (r5);
	\end{tikzpicture}
			\caption{Possible MinTSP for $p<0$ and MaxTSP solution for $0<p<1$ when $N=4$, which corresponds to the permutation cycles $(1,3,2,4)$ and $(1,3,4,2)$. Both of them have one crossing. One is obtained from the other by exchanging the two non-crossing matchings: the nested  $\{ \{1,4\},\{2,3\} \}$ with the parallel one  $\{ \{1,2\},\{3,4\} \}$}.
			\end{center}
\end{figure}

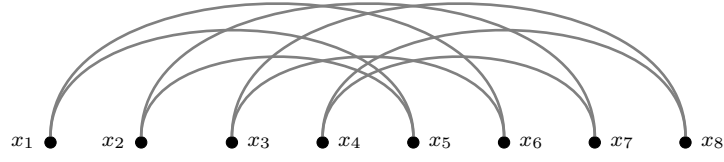
\begin{figure}[h]
	\begin{center}
	\begin{tikzpicture}[scale=0.6]
	\node[draw,circle,inner sep=1.5pt,fill=black,text=white,label=left:{\footnotesize $x_1$}] (r1) at (1,3) {};
	\node[draw,circle,inner sep=1.5pt,fill=black,text=white,label=left:{\footnotesize $x_2$}] (r2) at (3,3) {};
	\node[draw,circle,inner sep=1.5pt,fill=black,text=white,label=right:{\footnotesize $x_3$}] (r3) at (5,3) {};
	\node[draw,circle,inner sep=1.5pt,fill=black,text=white,label=right:{\footnotesize $x_4$}] (r4) at (7,3) {};
	\node[draw,circle,inner sep=1.5pt,fill=black,text=white,label=right:{\footnotesize $x_5$}] (r5) at (9,3) {};
	\node[draw,circle,inner sep=1.5pt,fill=black,text=white,label=right:{\footnotesize $x_6$}] (r6) at (11,3) {};
	\node[draw,circle,inner sep=1.5pt,fill=black,text=white,label=right:{\footnotesize $x_7$}] (r7) at (13,3) {};
	\node[draw,circle,inner sep=1.5pt,fill=black,text=white,label=right:{\footnotesize $x_8$}] (r8) at (15,3) {};
	\draw[line width=1pt,gray]  (r1) to[in=+90, out=+90] (r5);
	\draw[line width=1pt,gray]  (r5) to[in=+90, out=+90] (r2);
	\draw[line width=1pt,gray]  (r2) to[in=+90, out=+90] (r7);
	\draw[line width=1pt,gray]  (r7) to[in=+90, out=+90] (r4);
	\draw[line width=1pt,gray]  (r4) to[in=+90, out=+90] (r8);
	\draw[line width=1pt,gray]  (r8) to[in=+90, out=+90] (r3);
	\draw[line width=1pt,gray]  (r3) to[in=+90, out=+90] (r6);	
	\draw[line width=1pt,gray]  (r6) to[in=+90, out=+90] (r1);
	\end{tikzpicture}
			\caption{MinTSP for $p<0$ and MaxTSP solution for $0<p<1$ when $N=8$, which corresponds to the permutation cycle $(1,5,2,7,4,8,3,6)$.}
			\end{center}
\end{figure}

\section{The explicit solution for $p>1$}\label{solpg1}

The lowest possible value for $N$ for which the complete graph $\mathcal{K}_N$ allows a Hamiltonian cycle is $N=3$. In this case, there is only one Hamiltonian cycle, that is $(1,2,3)$.

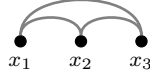
\begin{figure}[h]
	\begin{center}
	\begin{tikzpicture}[scale=0.4]
         \node[draw,circle,inner sep=1.5pt,fill=black,text=white,label=below:{\footnotesize $x_1$}] (p1) at (11,1.5) {};
	\node[draw,circle,inner sep=1.5pt,fill=black,text=white,label=below:{\footnotesize $x_2$}] (p2) at (13,1.5) {};
	\node[draw,circle,inner sep=1.5pt,fill=black,text=white,label=below:{\footnotesize $x_3$}] (p3) at (15,1.5) {};
				
	\draw[line width=1pt,gray]  (p1) to[in=+90, out=+90] (p2);
	\draw[line width=1pt,gray]  (p2) to[in=+90, out=+90] (p3);
	\draw[line width=1pt,gray]  (p1) to[in=+90, out=+90] (p3);
				
	\end{tikzpicture}
				\caption{The unique Hamiltonian cycle for $N=3$, which corresponds to the permutation cycle $(1,2,3)$.} 
	\end{center}
\end{figure}

For $N=4$ there are 3 Hamiltonian cycles corresponding to $A=(1,2,3,4), B=(1,3,4,2), C=(1,3,2,4)$.
In our one-dimensional Euclidean problem, the difference in cost between the cycles defined by $C$ and $A$ is the difference in cost between the crossing matching and the parallel one, while the difference in cost between the cycles defined by $C$ and $B$ is the difference in cost between the nested matching and the parallel one. It follows, from the Lemma applied to the vertices $\{1,2,3,4\}$, that $C$ provides the Hamiltonian cycle with $N=4$ with the highest cost.

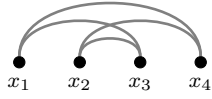
\begin{figure}[h]
	\begin{center}
	\begin{tikzpicture}[scale=0.4]
         \node[draw,circle,inner sep=1.5pt,fill=black,text=white,label=below:{\footnotesize $x_1$}] (p1) at (11,1.5) {};
	\node[draw,circle,inner sep=1.5pt,fill=black,text=white,label=below:{\footnotesize $x_2$}] (p2) at (13,1.5) {};
	\node[draw,circle,inner sep=1.5pt,fill=black,text=white,label=below:{\footnotesize $x_3$}] (p3) at (15,1.5) {};
	\node[draw,circle,inner sep=1.5pt,fill=black,text=white,label=below:{\footnotesize $x_4$}] (p4) at (17,1.5) {};
				
	\draw[line width=1pt,gray]  (p1) to[in=+90, out=+90] (p3);
	\draw[line width=1pt,gray]  (p1) to[in=+90, out=+90] (p4);
	\draw[line width=1pt,gray]  (p2) to[in=+90, out=+90] (p4);
	\draw[line width=1pt,gray]  (p2) to[in=+90, out=+90] (p3);
				
	\end{tikzpicture}
				\caption{{MaxTSP solution for $N=4$,  which corresponds to the permutation cycle $(1,3,2,4)$.} }
	\end{center}
\end{figure}

We are now in a position to prove that
\begin{thm}
For $p>1$, the cycle $\hat{h}$ defined, for $i\in[N]$, by
\begin{equation}
\hat{h}(i) =\begin{cases}
 \begin{cases}
	i  &\mbox{for odd } i\\ N-i+1 &\mbox{for even } i\\ \end{cases}  
  & N\, {\rm odd}\\[0.5cm]
 \begin{cases} i  &\mbox{for odd } i\leq N/2 \mbox{ and for even } i>N/2\\ N+1-i &\mbox{otherwise, } \end{cases}
  & N\, {\rm even}
 \end{cases} \label{solution}
\end{equation}
provides a Hamiltonian cycle with maximal cost.

It is unique up to the reverse order obtained by the exchange $i \leftrightarrow N-i+1$, which defines a mirror symmetry.
\end{thm}

\begin{proof}
The proof will be performed by induction on $N$, by assuming its validity for $N-2$ and by considering two different sequences, respectively, for the cases of odd and even $N$.

The lowest possible odd value for $N$ is $N=3$. In this case, there is only one Hamiltonian cycle, which is $(1,2,3)$ and is of the given form.

In the case of even $N$, the lowest possible value for $N$ is $N=4$.  We have seen that the permutation cycle with maximal cost is $(1,3,2,4)$, in agreement with~(\ref{solution}).

First, we shall prove that the edge $e_{1N}$ belongs to the optimal Hamiltonian cycle. By contradiction, assume that the optimal cycle is of the form
$$(1, a, \dots, c, N, d, \ldots, b)$$
with $b\neq c$ because it is a cycle.
Consider the following change: delete in the Hamiltonian cycle the edges $e_{c ,N}$ and $e_{b, 1}$ and add $e_{1,N}$ and $e_{b,c}$ instead. This is a valid cycle. Now $1<b$ and $c<N$. Now consider the quadrupole $\{1, \min(b,c), \max(b,c), N\}$. If $b<c$, the two removed edges are parallel; if $b>c$, they are crossed. The edges $e_{1,n}, e_{b,c}$ are exactly the nested ones. This change increases the cost, contradicting the assumption that the cycle is optimal.

Since $e_{1,N}$ belongs to the optimal Hamiltonian cycle, so do $e_{1,N-1}$ and $e_{2,N}$.
Indeed, if we remove the edge $e_{1,N}$ we are left with the Hamiltonian path $P$
$$ 1=P(0), P(1), \ldots, P(N-1)=N .$$
As the cost of the edge $e_{1N}$, which is $\phi(x_N-x_1)$, is fixed, maximizing the cost of the Hamiltonian cycles is equivalent to maximizing the cost of the Hamiltonian paths from the vertex 1 to the vertex $N$. Assume, by contradiction, that $P(1)\neq N-1$. Then $P(i)=N-1$ for some $2\leq i \leq N-2$. Reverse the sub-path $P(1),\dots,P(i)$, replace with the new edges $e_{1,N-1}$ and $e_{P(1), P(i+1)}$ the old edges $e_{1,P(1)}$ and $e_{N-1, P(i+1)}$,  when $i\leq N-3$,  otherwise with $e_{N-1,N}$ in the case in which $i=N-2$, that is $N-1$ was a neighbour of $N$. 
In the case $2\leq i \leq N-3$, we can apply our Lemma to the quadrupole $\{1, \min(P(1),P(i+1)), \max(P(1),P(i+1)), N-1\}$: the old edges form the parallel or crossing matchings, the new ones the nested matching, which makes a higher cost. In the last case, that is $i=N-2$, by the choice of the quadrupole $\{1, P(1), N-1,N\}$ in our Lemma, we still get a higher cost, as the old edges form a parallel matching while the new ones form a crossing matching. And this is a contradiction.
By the mirror symmetry $i \leftrightarrow N-i+1$,  the same argument shows that the other neighbour of $N$ is the vertex $2$.

We are now ready for the final step: the reduction in $N$. As we have seen that the Hamiltonian cycle defined by $\hat{h}$ contains the edges $e_{1,N}, e_{1,N-1}, e_{2,N}$, delete these three edges and the  vertices $1$ and $N$. Close the remaining path with the new edge $e_{2,N-1}$ and this defines a new Hamiltonian cycle on $N-2$ vertices which corresponds to the permutation cycle $h'$ in $\mathcal{S}_{N-2}$ which acts on $\{2, \ldots,N-1\}$. 
The costs are related by
$$
E(\hat{h}) = E(h') +\left[ \phi(x_N-x_1) + \phi(x_{N-1}-x_1) +\phi(x_N-x_2) - \phi(x_{N-1}-x_2)\right] .
$$
As the contribution in square brackets is fixed, the optimal cycle on $N$ vertices in $[N]$ is equivalent to the optimal cycle on the $N-2$ vertices $\{2, \ldots,N-1\}$, which proves the reduction step and completes the proof.
\end{proof}

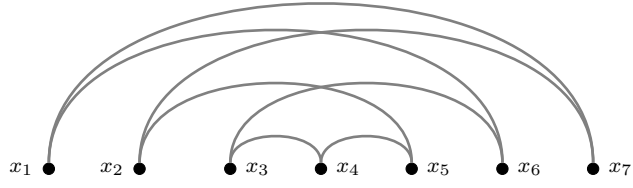
\begin{figure}[h!]
		\begin{center}
			\begin{tikzpicture}[scale=0.6]
			\node[draw,circle,inner sep=1.5pt,fill=black,text=white,label=left:{\footnotesize $x_1$}] (r1) at (1,3) {};
			\node[draw,circle,inner sep=1.5pt,fill=black,text=white,label=left:{\footnotesize $x_2$}] (r2) at (3,3) {};
			\node[draw,circle,inner sep=1.5pt,fill=black,text=white,label=right:{\footnotesize $x_3$}] (r3) at (5,3) {};
			\node[draw,circle,inner sep=1.5pt,fill=black,text=white,label=right:{\footnotesize $x_4$}] (r4) at (7,3) {};
			\node[draw,circle,inner sep=1.5pt,fill=black,text=white,label=right:{\footnotesize $x_5$}] (r5) at (9,3) {};
			\node[draw,circle,inner sep=1.5pt,fill=black,text=white,label=right:{\footnotesize $x_6$}] (r6) at (11,3) {};
			\node[draw,circle,inner sep=1.5pt,fill=black,text=white,label=right:{\footnotesize $x_7$}] (r7) at (13,3) {};
			
			\draw[line width=1pt,gray]  (r1) to[in=+90, out=+90] (r6);
			\draw[line width=1pt,gray]  (r6) to[in=+90, out=+90] (r3);
			\draw[line width=1pt,gray]  (r3) to[in=+90, out=+90] (r4);
			\draw[line width=1pt,gray]  (r4) to[in=+90, out=+90] (r5);
			\draw[line width=1pt,gray]  (r2) to[in=+90, out=+90] (r5);
			\draw[line width=1pt,gray]  (r2) to[in=+90, out=+90] (r7);
			\draw[line width=1pt,gray]  (r7) to[in=+90, out=+90] (r1);	
			\end{tikzpicture}
			\caption{MaxTSP solution when $p>1$  for $N=7$, which corresponds to the permutation cycle $(1,6,3,4,5,2,7)$.}
		\end{center}
\end{figure}

\begin{figure}[h!]
	\begin{center}
	\begin{tikzpicture}[scale=0.6]
	\node[draw,circle,inner sep=1.5pt,fill=black,text=white,label=left:{\footnotesize $x_1$}] (r1) at (1,3) {};
	\node[draw,circle,inner sep=1.5pt,fill=black,text=white,label=left:{\footnotesize $x_2$}] (r2) at (3,3) {};
	\node[draw,circle,inner sep=1.5pt,fill=black,text=white,label=right:{\footnotesize $x_3$}] (r3) at (5,3) {};
	\node[draw,circle,inner sep=1.5pt,fill=black,text=white,label=right:{\footnotesize $x_4$}] (r4) at (7,3) {};
	\node[draw,circle,inner sep=1.5pt,fill=black,text=white,label=right:{\footnotesize $x_5$}] (r5) at (9,3) {};
	\node[draw,circle,inner sep=1.5pt,fill=black,text=white,label=right:{\footnotesize $x_6$}] (r6) at (11,3) {};
	\node[draw,circle,inner sep=1.5pt,fill=black,text=white,label=right:{\footnotesize $x_7$}] (r7) at (13,3) {};
	\node[draw,circle,inner sep=1.5pt,fill=black,text=white,label=right:{\footnotesize $x_8$}] (r8) at (15,3) {};
	\draw[line width=1pt,gray]  (r1) to[in=+90, out=+90] (r8);
	\draw[line width=1pt,gray]  (r2) to[in=+90, out=+90] (r8);
	\draw[line width=1pt,gray]  (r1) to[in=+90, out=+90] (r7);
	\draw[line width=1pt,gray]  (r3) to[in=+90, out=+90] (r7);
	\draw[line width=1pt,gray]  (r3) to[in=+90, out=+90] (r5);
	\draw[line width=1pt,gray]  (r4) to[in=+90, out=+90] (r5);
	\draw[line width=1pt,gray]  (r6) to[in=+90, out=+90] (r4);	
	\draw[line width=1pt,gray]  (r6) to[in=+90, out=+90] (r2);
	\end{tikzpicture}
	\caption{MaxTSP solution when $p>1$ for $N=8$, which corresponds to the permutation cycle $(1,7,3,5,4,6,2,8)$.}
\end{center}
\end{figure}
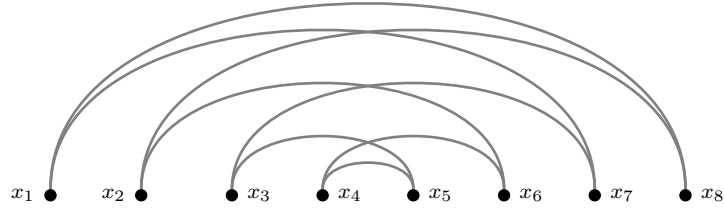

For comparison, recall that the MinTSP is instead provided by the permutation cycle~\cite{Caracciolo:178}
\be
\hat{h}(i) = \begin{cases}
2 i -1 & \mbox{for \,} i \leq \frac{N+1}{2}\\
2N-2i+2 & \mbox{for \,} i >  \frac{N+1}{2}.
\end{cases}
\ee

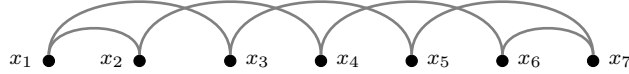
\begin{figure}[h!]
		\begin{center}
			\begin{tikzpicture}[scale=0.6]
			\node[draw,circle,inner sep=1.5pt,fill=black,text=white,label=left:{\footnotesize $x_1$}] (r1) at (1,3) {};
			\node[draw,circle,inner sep=1.5pt,fill=black,text=white,label=left:{\footnotesize $x_2$}] (r2) at (3,3) {};
			\node[draw,circle,inner sep=1.5pt,fill=black,text=white,label=right:{\footnotesize $x_3$}] (r3) at (5,3) {};
			\node[draw,circle,inner sep=1.5pt,fill=black,text=white,label=right:{\footnotesize $x_4$}] (r4) at (7,3) {};
			\node[draw,circle,inner sep=1.5pt,fill=black,text=white,label=right:{\footnotesize $x_5$}] (r5) at (9,3) {};
			\node[draw,circle,inner sep=1.5pt,fill=black,text=white,label=right:{\footnotesize $x_6$}] (r6) at (11,3) {};
			\node[draw,circle,inner sep=1.5pt,fill=black,text=white,label=right:{\footnotesize $x_7$}] (r7) at (13,3) {};
			
			\draw[line width=1pt,gray]  (r1) to[in=+90, out=+90] (r3);
			\draw[line width=1pt,gray]  (r3) to[in=+90, out=+90] (r5);
			\draw[line width=1pt,gray]  (r5) to[in=+90, out=+90] (r7);
			\draw[line width=1pt,gray]  (r7) to[in=+90, out=+90] (r6);
			\draw[line width=1pt,gray]  (r6) to[in=+90, out=+90] (r4);
			\draw[line width=1pt,gray]  (r4) to[in=+90, out=+90] (r2);
			\draw[line width=1pt,gray]  (r2) to[in=+90, out=+90] (r1);	
			\end{tikzpicture}
			\caption{MinTSP solution when $p>1$  for $N=7$, which corresponds to the permutation cycle $(1,3,5,7,6,4,2)$ or its reverse $(1,2,4,6,7,5,3)$.}
		\end{center}
\end{figure}

\begin{figure}[h!]
	\begin{center}
	\begin{tikzpicture}[scale=0.6]
	\node[draw,circle,inner sep=1.5pt,fill=black,text=white,label=left:{\footnotesize $x_1$}] (r1) at (1,3) {};
	\node[draw,circle,inner sep=1.5pt,fill=black,text=white,label=left:{\footnotesize $x_2$}] (r2) at (3,3) {};
	\node[draw,circle,inner sep=1.5pt,fill=black,text=white,label=right:{\footnotesize $x_3$}] (r3) at (5,3) {};
	\node[draw,circle,inner sep=1.5pt,fill=black,text=white,label=right:{\footnotesize $x_4$}] (r4) at (7,3) {};
	\node[draw,circle,inner sep=1.5pt,fill=black,text=white,label=right:{\footnotesize $x_5$}] (r5) at (9,3) {};
	\node[draw,circle,inner sep=1.5pt,fill=black,text=white,label=right:{\footnotesize $x_6$}] (r6) at (11,3) {};
	\node[draw,circle,inner sep=1.5pt,fill=black,text=white,label=right:{\footnotesize $x_7$}] (r7) at (13,3) {};
	\node[draw,circle,inner sep=1.5pt,fill=black,text=white,label=right:{\footnotesize $x_8$}] (r8) at (15,3) {};
	\draw[line width=1pt,gray]  (r1) to[in=+90, out=+90] (r3);
	\draw[line width=1pt,gray]  (r3) to[in=+90, out=+90] (r5);
	\draw[line width=1pt,gray]  (r5) to[in=+90, out=+90] (r7);
	\draw[line width=1pt,gray]  (r7) to[in=+90, out=+90] (r8);
	\draw[line width=1pt,gray]  (r8) to[in=+90, out=+90] (r6);
	\draw[line width=1pt,gray]  (r6) to[in=+90, out=+90] (r4);
	\draw[line width=1pt,gray]  (r4) to[in=+90, out=+90] (r2);	
	\draw[line width=1pt,gray]  (r2) to[in=+90, out=+90] (r1);
	\end{tikzpicture}
	\caption{MinTSP solution when $p>1$  for $N=8$, which corresponds to the permutation cycle $(1,3,5,7,8,6,4,2)$ or its reverse  $(1,2,4,6,8,7,5,3)$.}
\end{center}
\end{figure}
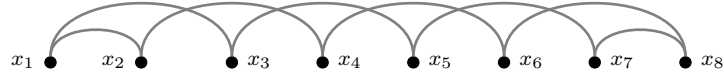

\section{Average optimal cost.}\label{AOC}

As explained before, our main interest is the average properties of random combinatorial problems. We shall concentrate on the case in which the $N$ random points $\{x_1,\ldots,x_N\}$ are chosen independently and uniformly in the interval $[0,1]$, exactly as it was done for the MinTSP in~\cite{Caracciolo:178}, which offers a direct comparison.

Once the $N$ random points are ordered by increasing position, their distribution is given by 
\be
p(x_1,\ldots,x_N) = N! \prod_{i=0}^N \theta(x_{i+1}-x_i) 
\ee
where $x_0=0$ and $x_{N+1}=1$. 
It follows that the probability for finding  the $l$-th point in $[x,x+dx]$ is given by
\begin{equation}
p_l(x) dx = \frac{\Gamma(N+1)}{\Gamma(l) \, \Gamma(N-l+1)} \, x^{l-1} (1-x)^{N-l} dx
\end{equation}
while the probability for finding the $l$-th point in $[x,x+dx]$ and the $s$-th point in $[y,y+dy]$ is given, for $s>l$  by
\begin{equation}
\begin{split}
p_{l,s}(x, y) dx dy  & = \frac{\Gamma(N+1)}{\Gamma(l) \, \Gamma(s-l) \, \Gamma(N-s+1)} \, x^{l-1} \\
& \quad \times (y-x)^{s-l-1} (1-y)^{N-s}\, \theta(y-x) dx dy \,,
\end{split}
\end{equation}
For the evaluation of the average cost, when the cost function is given by $\phi(x)=x^p, \,  p\in \mathbb{R}$, we shall only need the expectation for $k<N$
\begin{equation}
\mathbb{E} \left[ (x_{l+k}- x_l)^\alpha \right] = \int \,dx\, dy\, (y-x)^\alpha\, p_{l, \, l+k}(x, y) = \frac{\Gamma(N+1)\, \Gamma(k+\alpha)}{\Gamma(N+\alpha+1)\, \Gamma(k)   } \label{cost1}
\end{equation}
which does not depend on $l$, because of the invariance under translations. Remark that the integral is convergent only for $k>-\alpha$. 

We shall now proceed with the evaluation of the expectation value of the cost of the MaxTSP, which will be a different function of $N$ and $p$ in the different regions where we have determined the Hamiltonian cycle $\hat{h}_{\M}$.

\subsection{The region $\mathbf{p>1}$}
In the case in which $N$ is even, as
\be
\begin{split}
 E(\hat{h}_{\M})  =  &\phi(x_N-x_1) + \phi(x_{\frac{N}{2}+1}-x_{\frac{N}{2}}) \\ & + \sum_{k=1}^{\frac{N}{2}-1} \left[ \phi(x_{\frac{N}{2}+k} -x_{\frac{N}{2}-k}) + \phi(x_{\frac{N}{2}+1+k} -x_{\frac{N}{2}+1-k})\right]
\end{split}
\ee
we recover that		
\begin{equation}
 \mathbb{E} \left[ E(\hat{h}_{\M}) \right] =  \frac{\Gamma(N+1)}{\Gamma(N+p+1)} \left[  \frac{\Gamma(N+p-1)}{\Gamma(N-1)} +
 \frac{\Gamma(p+1)}{\Gamma(1)} + 2 \sum_{k=1}^{\frac{N}{2}-1}               \frac{\Gamma(2 k+p)}{\Gamma(2k)} \right] .
 \end{equation}		
The sum is conveniently split as follows
\be
2 \sum_{k=1}^{\frac{N}{2}-1} \frac{\Gamma(2 k+p)}{\Gamma(2k)} = \sum_{i=1}^{N-2} \frac{\Gamma(i+p)}{\Gamma(i)} +
 \sum_{i=1}^{N-2} (-1)^i \frac{\Gamma(i+p)}{\Gamma(i)} 
\ee	
and
\be
\sum_{i=1}^{N-2} \frac{\Gamma(i+p)}{\Gamma(i)} = \frac{\Gamma(N+p-1)}{(p+1) \Gamma(N-2)}
\ee
while 
\begin{align}
 \sum_{i=1}^{N-2} (-1)^i \frac{\Gamma(i+p)}{\Gamma(i)} & = \sum_{i=1}^{\infty} (-1)^i \frac{\Gamma(i+p)}{\Gamma(i)} -\sum_{i=N-1}^{\infty} (-1)^i \frac{\Gamma(i+p)}{\Gamma(i)} \nonumber \\
 & = -\frac{\Gamma(p+1)}{2^{p+1}} + (-1)^N \frac{\Gamma(N+p-1) {}_2F_1(1,N+p-1;N-1;-1)}{\Gamma(N-1) }
 \end{align}
where $ {}_2F_1$ is the (1,2)-hypergeometric function.
In conclusion, when $N$ is even
\be
\begin{split}
 \mathbb{E} \left[ E(\hat{h}_{\M}) \right]  = & \frac{N (N-1)}{(p+1)(N+p)}+ \frac{\Gamma(N+1) \Gamma(p+1)}{\Gamma(N+p+1)} \left(1 - \frac{1}{2^{p+1}} \right)\\
& + \frac{\Gamma(N+p-1) \Gamma(N+1)}{\Gamma(N+p+1) \Gamma(N-1)}\, {}_2F_1(1,N+p-1;N-1;-1).
 \end{split}	 \label{aepg1Ne}
\ee
Also, when $N$ is odd, we can apply the same strategy. 
As
\be
\begin{split}
 E(\hat{h}_{\M})  =  &\phi(x_N-x_1) + 
 \\ & + \sum_{k=1}^{\frac{N-1}{2}} \left[ \phi(x_{\frac{N+1}{2}+k} -x_{\frac{N+1}{2}-k+1}) + \phi(x_{\frac{N+1}{2}-1+k} -x_{\frac{N+1}{2}-k})\right]
\end{split}
\ee
we recover that		
\begin{equation}
 \mathbb{E} \left[ E(\hat{h}_{\M}) \right] =  \frac{\Gamma(N+1)}{\Gamma(N+p+1)} \left[  \frac{\Gamma(N+p-1)}{\Gamma(N-1)} 
  + 2 \sum_{k=1}^{\frac{N-1}{2}}               \frac{\Gamma(2 k-1+p)}{\Gamma(2k-1)} \right] .
 \end{equation}		
Now 
\be
2 \sum_{k=1}^{\frac{N-1}{2}} \frac{\Gamma(2 k-1+p)}{\Gamma(2k-1)} = \sum_{i=1}^{N-2} \frac{\Gamma(i+p)}{\Gamma(i)} -
 \sum_{i=1}^{N-2} (-1)^i \frac{\Gamma(i+p)}{\Gamma(i)} 
\ee	
In conclusion, when $N$ is odd 
\be
\begin{split}
 \mathbb{E} \left[ E(\hat{h}_{\M}) \right]  = & \frac{N (N-1)}{(p+1)(N+p)}+ \frac{\Gamma(N+1) \Gamma(p+1)}{2^{p+1}\Gamma(N+p+1)} \\
& + \frac{\Gamma(N+p-1) \Gamma(N+1)}{\Gamma(N+p+1) \Gamma(N-1)}\, {}_2F_1(1,N+p-1;N-1;-1).\label{aepg1No}
\end{split}
\ee
For integer values of $p$ and $N>2$, the previous expressions for the average values of the cost become rational functions of $N$ (see Appendix~\ref{AppA}).	
For example, for $p=2$ we get
\be
\mathbb{E} \left[ E(\hat{h}_{\M}) \right]  = \frac{N}{3} -\frac{1}{2} - 
\begin{cases}
\frac{N-3}{2 (N+1) (N+2)} & \mbox{for odd\, } N\\
\frac{N-6}{2 (N+1) (N+2)} & \mbox{for even\, } N.
\end{cases}
\ee
From the explicit expressions for the number of points $N$, both even and odd, the asymptotic behaviour of the average cost for large  $N$  is derived. The result is
 \be
 \mathbb{E} \left[ E(\hat{h}_{\M}) \right]  = \frac{N}{p+1} -\frac{1}{2} + \mathcal{O}\left(\frac{1}{N}\right).
 \ee
 
Let us compare these costs with the one for the MinTSP, when
\be
E(\hat{h}_{\m}) = \phi(x_2-x_1) +\phi(x_N-x_{N-1}) +\sum_{k=1}^{N-2} \phi(x_{k+2}-x_k)
\ee
so that
\be
\mathbb{E} \left[ E(\hat{h}_{\m}) \right]  = \frac{\Gamma(N+1) \Gamma(p+1)}{\Gamma(N+p+1)} \left[ 2 + (N-2) (p+1)\right] 
\ee
which for large $N$ becomes
\be
 \mathbb{E} \left[ E(\hat{h}_{\m}) \right] = N^{-p} \left[ N\,  \Gamma(p+2) - \frac{p (p^2 + 2 p +5) \Gamma(p+1)}{2} + \mathcal{O}\left(\frac{1}{N}\right) \right] .
\ee 
We see that, while in the MinTSP each step gives on average a contribution to the cost of order $N^{-p}$, in the MaxTSP it is instead of order 1.

\subsection{The region $\mathbf {0<p<1}$}
When $N$ is odd, the cost of the optimal solution is
\be
E(\hat{h}_{\M})  = \phi(x_{N}-x_{\frac{N+1}{2}}) + \sum_{k=1}^{\frac{N-1}{2}}\left[ \phi(x_{k+\frac{N-1}{2}}-x_k) + \phi(x_{k+\frac{N+1}{2}}-x_k) \right]
\ee
which provides the average cost
\begin{equation}
\mathbb{E} \left[ E(\hat{h}_{\M}) \right] = \frac{\Gamma(N+1)}{2\,\Gamma(N+p+1)} \left[(N-1) \frac{\Gamma\left(\frac{N+1}{2}+p\right)}{\Gamma \left( \frac{N+1}{2} \right)} + (N+1) \frac{\Gamma\left(\frac{N-1}{2}+p\right)}{\Gamma \left( \frac{N-1}{2} \right)}\right] \,. \label{aepl1No}
\end{equation}	
In the asymptotic regime of large number of points $N$
\be
 \mathbb{E} \left[ E(\hat{h}_{\M}) \right]  = 2^{-p} N  - 2^{-p-1}  p\, (3- p)+ \mathcal{O}\left(\frac{1}{N}\right).\label{asdecodd}
\ee
When $N$ is even, we have remarked that there is not a unique optimal solution. There are $\frac{N}{2}$ of them. Each of them provides the MaxTSP with particular choices of the positions of the $N$ points. By taking the permutation cycles $\sigma_k(\bar{h})$ with $k\in[N]$, each Hamiltonian cycle appears twice. Therefore we have, in this case, that the average costs of the Hamiltonian cycles corresponding to this set of permutations are only lower bounds on the average optimal cost.

 The cost for the permutation $\bar{h} = \sigma_1(\bar{h})$ is
\be
\begin{split}
E(\bar{h})  = & \phi(x_{\frac{N}{2}+1}-x_1) + \phi(x_N - x_{\frac{N}{2}})+\\
& +\sum_{k=2}^{\frac{N}{2}} \left[ \phi(x_{\frac{N}{2}+k-1}-x_k) + \phi(x_{\frac{N}{2}+k} - x_{k-1})\right]
\end{split}
\ee	
which has the average
\begin{equation}
\begin{split}
\mathbb{E}\left[E(\bar{h} )\right]= & \frac{\Gamma(N+1)}{\Gamma(N+p+1)} \left[ \left(\frac{N}{2}-1\right) \frac{\Gamma\left( \frac{N}{2} + p-1 \right)}{\Gamma \left( \frac{N}{2}-1 \right)} \right. \\
			& \left. + \left( \frac{N}{2}-1 \right) \frac{\Gamma \left( \frac{N}{2} + p + 1 \right)}{\Gamma \left( \frac{N}{2} +1 \right)} + 2 \frac{\Gamma\left( \frac{N}{2}+p \right)}{\Gamma \left( \frac{N}{2} \right)}\right].
\end{split}
\label{hbar}
\end{equation}
In the asymptotic regime of a large number of points $N$
\be
 \mathbb{E} \left[ E(\bar{h} ) \right]  = 2^{-p} N  - 2^{-1-p} p\, (3-  p)+ \mathcal{O}\left(\frac{1}{N}\right) 
\ee
which agrees with the expansion for odd $N$ in (\ref{asdecodd}).
All the other permutations, except for the one corresponding to $\hat{h}$ in reverse order, have the same average cost
\begin{equation}
\begin{split}
\mathbb{E}\left[E(h^*)\right]= & \frac{\Gamma(N+1)}{\Gamma(N+p+1)} \left[ \frac{N}{2}  \frac{\Gamma\left( \frac{N}{2} + p-1 \right)}{\Gamma \left( \frac{N}{2}-1 \right)} \right. \\
			& \left. + \left( \frac{N}{2}-2 \right) \frac{\Gamma \left( \frac{N}{2} + p + 1 \right)}{\Gamma \left( \frac{N}{2} +1 \right)} + 2 \frac{\Gamma\left( \frac{N}{2}+p \right)}{\Gamma \left( \frac{N}{2} \right)}\right].
\end{split}
\label{hstar}
\end{equation}
It soon follows that $\bar{h}_{\M}$ has a higher cost, as
\be
0<  \mathbb{E} \left[ E(\bar{h}_{\M} ) \right] - \mathbb{E}\left[E(h^*)\right] =  \frac{\Gamma(N+1)}{\Gamma(N+p+1)} \left[ \frac{\Gamma \left( \frac{N}{2} + p + 1 \right)}{\Gamma \left( \frac{N}{2} +1 \right)} - \frac{\Gamma \left( \frac{N}{2} + p - 1 \right)}{\Gamma \left( \frac{N}{2} -1 \right)} \right].
\ee
So we get the bounds
\be
\mathbb{E} \left[ E(\hat{h}_{\M} ) \right] > \mathbb{E} \left[ E(\bar{h} ) \right] > \mathbb{E}\left[E(h^*)\right] .
\ee
In the asymptotic regime of large $N$, we see that
\be
\mathbb{E} \left[ E(\bar{h} ) \right] - \mathbb{E}\left[E(h^*)\right] = \mathcal{O}\left(\frac{1}{N}\right).
\ee
So we expect that also
\be
\mathbb{E} \left[ E(\hat{h}_{\M} ) \right] - \mathbb{E}\left[E(\bar{h})\right] = \mathcal{O}\left(\frac{1}{N}\right).
\ee

It is interesting to notice that, for both $N$ even and $N$ odd
\be
\lim_{p\to 1^+} \lim_{N\to \infty} \mathbb{E} \left[ E(\hat{h}_{\M}) - \frac{N}{p+1} \right] =  -\frac{1}{2}
\ee
and also for $N$ odd
\be
\lim_{p\to 1^-} \lim_{N\to \infty} \mathbb{E} \left[ E(\hat{h}_{\M} )- \frac{N}{2^p}  \right] =  -\frac{1}{2}.
\ee
But also for $N$ even 
\be
\lim_{p\to 1^-} \lim_{N\to \infty} \mathbb{E} \left[ E(\bar{h}) - \frac{N}{2^p} \right] =  -\frac{1}{2}.
\ee

\subsection{The region $\mathbf{p<0}$}
The cost of the optimal solution is, in this case
\be
E(\hat{h}_{\M})  = \phi(x_{N}-x_1) +\sum_{k=1}^{N-1}  \phi(x_{k+1}-x_k) 
\ee
and therefore, the average cost is
\begin{equation}
 \mathbb{E} \left[ E(\hat{h}_{\M}) \right]  = \frac{\Gamma(N+1)}{\Gamma(N+p+1)} \left[ \frac{\Gamma(N + p - 1)}{\Gamma(N-1)} +(N-1) \, \Gamma(p+1) \right]  \label{aepl0}
\end{equation}
at least for $-1<p<0$, otherwise it is divergent.

In the asymptotic regime of large number of points $N$
\be
 \begin{aligned}
 \mathbb{E} \left[ E(\hat{h}_{\M}) \right]  =
&\frac{\Gamma(p+1)}{N^p}\left[ N-1-\frac{p(p+1)}{2}+\mathcal{O}\left(\frac{1}{N}\right)\right]\\
&\qquad +1-\frac{2p}{N}+\mathcal{O}\left(\frac{1}{N^2}\right)
\end{aligned}
 \ee
where the leading contribution comes from the $N-1$ steps from a point to its neighbour, which have a cost of order $N^{-p}$.

\section{Conclusions}\label{end}

We have considered the MaxTSP for $N$ points on the unit interval $[0,1]$ for various choices of the cost function $f(x,y)$ for the edge $\{x,y\}$, with $x,y\in[0,1]$, when $f(x,y) = \phi(\mbox{dist}(x,y))$, where $\mbox{dist}(x,y)=|y-x|$ is the Euclidean distance between the two points. We focused on the particular form $\phi(x)=x^p$ for different values of the parameter $p$.

The problem is simple enough to allow us to determine exactly the MaxTSP for every value of the number of points, with one exception: in the region $0<p<1$ and for an even number of points $N$, there are, in general, $\frac{N}{2}$ different Hamiltonian cycles that achieve the maximum cost for different realizations of the disorder.

We can, therefore, in the case in which the disorder for the choice of the position of the points is the flat distribution,  determine the average cost for a random Euclidean combinatorial problem in the presence of strong correlations among the different edges.

At finite $N$ we derive, in general, different explicit expressions for the average optimal cost for the cases of $N$ even or odd, always with the exception of the region $0<p<1$ and even number of points $N$. See formulas~(\ref{aepg1No}) and~(\ref{aepg1Ne}), respectively for the cases $N$ odd and even, in the region $p>1$;  (\ref{aepl1No}) for the case $N$ odd, in the region $0<p<1$; (\ref{aepl0}) in the region $-1<p<0$. For $p\leq-1$ it is divergent.
They are analytic functions of the parameters $N$ and $p$. For integer $p>1$, they are rational functions of $N$.
In the asymptotic regime of large number of points, the expressions for even and odd $N$ coincide in the leading term and in the first sub-leading one, which we summarize here
\begin{equation*}
\mathbb{E}\left[E(\hat{h}_{\M})\right] =
\begin{cases}
\displaystyle
\frac{N}{p+1}-\frac{1}{2}+\mathcal{O}\left(\frac{1}{N}\right) & \text{when } p>1,\\[1.2em]
\displaystyle
\frac{N}{2^p}-\frac{p(3-p)}{2^{p+1}}+\mathcal{O}\left(\frac{1}{N}\right) & \text{when } 0<p<1,\\[1.2em]
\displaystyle
\begin{aligned}
&\frac{\Gamma(p+1)}{N^p}\left[ N-1-\frac{p(p+1)}{2}+\mathcal{O}\left(\frac{1}{N}\right)\right]\\
&\qquad +1-\frac{2p}{N}+\mathcal{O}\left(\frac{1}{N^2}\right)
\end{aligned} 
& \text{when } -1<p<0
\end{cases}
\end{equation*}
for the convenience of the readers.

\appendix
\section{More about our hypergeometric functions}\label{AppA}

By Euler's transformation for the hypergeometric function 
 \be
 {}_2F_1(a,b;c;z) = (1-z)^{c-a-b} {}_2F_1(c-a,c-b;c;z)
  \ee
taking $a=1, b=N+p-1, c=N-1, z=-1$, we obtain
\be
 {}_2F_1(1,N+p-1;N-1;-1) = \,\frac{1}{2^{p+1}} {}_2F_1(N-2,-p;N-1;-1) .
 \ee
 Since $-p<0$, when $p$ is a natural number, and $N>2$, the transformed hypergeometric series terminates. Indeed
 \be
  \begin{split}
  {}_2F_1(1,N+p-1;N-1;-1) = & \,\frac{1}{2^{p+1}} \sum_{k=0}^p \binom{p}{k} \frac{(N-2)_k}{(N-1)_k}\\
   = & \,\frac{N-2}{2^{p+1} }\sum_{k=0}^p \binom{p}{k} \frac{1}{N+k-2}.
 \end{split}
  \ee
  where $(a)_k$ is the rising factorial. We conclude that the average cost of the optimal solution, for integer $p>1$ becomes a rational function of $N$.

It soon follows, from this representation, that, for large $N$
\be
{}_2F_1(1,N+p-1;N-1;-1) = \frac{1}{2} - \frac{p}{4 N} +\mathcal{O}\left(\frac{1}{N^2}\right)
\ee
and that, for example, for $p=2$
\be
{}_2F_1(1,N+1;N-1;-1) = \frac{2 N^2 - 4 N +1}{4 N (N-1)} .
\ee

\section{Large $N$ behavior}\label{AppB}
We report here the leading order term in the thermodynamic limit of the
average optimal cost, for independent uniformly distributed points in $[0,1]$
and fixed $p$.
\subsection{\bf $p>1$, $N$ even} 
We need to find the leading order of the term
$\sum_{k=1}^{N/2-1}\Gamma(2k+p)/\Gamma(2k)$. Let us introduce the generating function
\begin{equation}
 f(z)=\sum_{M\geq0}\left(\sum_{k=1}^{M-1}
 \frac{\Gamma(2k+p)}{\Gamma(2k)}\right)z^M,
\end{equation}
where empty sums are zero. With $M=N/2$, we are interested in the $M$-th
coefficient in the expansion of the generating function. We can write
\begin{equation}
\begin{split}
 f(z)&=\sum_{k=1}^{\infty}\sum_{M=k+1}^{\infty}
 \frac{\Gamma(2k+p)}{\Gamma(2k)}z^M\\
 &=\frac{z}{1-z}\sum_{k=1}^{\infty}
 \frac{\Gamma(2k+p)}{\Gamma(2k)}z^k,
\end{split}
\end{equation}
where we set $M-k-1=j$. Performing the sum gives
\begin{equation}
 f(z)=\frac{\Gamma(p+1)z^{3/2}}{2(1-z)^{p+2}}
 \left[(1+\sqrt z)^{p+1}-(1-\sqrt z)^{p+1}\right].
\end{equation}
Despite the square roots, this expression is analytic at $z=0$: after
setting $z=t^2$, it is an even analytic function of $t$.
Near the dominant singularity $z=1$,
\begin{equation}
 f(z)=2^p\Gamma(p+1)(1-z)^{-p-2}
 +\mathcal O\bigl((1-z)^{-p-1}\bigr).
\end{equation}
We extract the coefficient of this singular term using
\cite{flajolet2009analytic}
\begin{equation}
 [z^M](1-z)^{-p-2}
 =\frac{\Gamma(M+p+2)}{\Gamma(p+2)\Gamma(M+1)}
 =\frac{M^{p+1}}{\Gamma(p+2)}+\mathcal O(M^p).
\label{eq:leading-transfer}
\end{equation}
Thus
\begin{equation}
 [z^M]f(z)=\frac{2^p M^{p+1}}{p+1}+\mathcal O(M^p)
 =\frac{N^{p+1}}{2(p+1)}+\mathcal O(N^p).
\end{equation}
We are now ready to determine the leading order of the average optimal cost.
The gamma ratios satisfy
\begin{align}
 \frac{\Gamma(N+1)}{\Gamma(N+p+1)}
 &=N^{-p}\left[1-\frac{p^2+p}{2N}+\mathcal O(N^{-2})\right],\\
 \frac{\Gamma(N+p-1)}{\Gamma(N-1)}
 &=N^p\left[1+\frac{p^2-3p}{2N}+\mathcal O(N^{-2})\right].
\end{align}
In the expression for the cost, the sum occurs with a factor $2$, while
the remaining terms give only $\mathcal O(1)$ after multiplication by
$\Gamma(N+1)/\Gamma(N+p+1)$. 

\subsection{\bf $p>1$, $N$ odd} 
The same procedure applies to
$\sum_{k=1}^{(N-3)/2}\Gamma(2k+1+p)/\Gamma(2k+1)$. We define
\begin{equation}
 f(z)=\sum_{M\geq0}\left(\sum_{k=1}^{M}
 \frac{\Gamma(2k+1+p)}{\Gamma(2k+1)}\right)z^M,
 \qquad M=\frac{N-3}{2}.
\end{equation}
Performing the sum gives
\begin{equation}
 f(z)=\frac{\Gamma(p+1)}{2(1-z)^{p+2}}
 \left[(1-\sqrt z)^{p+1}+(1+\sqrt z)^{p+1}
 -2(1-z)^{p+1}\right].
\end{equation}
The square roots again cancel in the expansion at the origin. Near $z=1$,
the leading singular term is the same as in the even case, so
\begin{equation}
 [z^M]f(z)=\frac{2^pM^{p+1}}{p+1}+\mathcal O(M^p)
 =\frac{N^{p+1}}{2(p+1)}+\mathcal O(N^p).
\end{equation}

\subsection{\bf $0<p<1$} 
For $N=2M+1$, the exact average cost can be written as
\begin{equation}
 \mathbb{E}\left[E(\hat{h}_{\M})\right]
 =\frac{\Gamma(N+1)}{\Gamma(N+p+1)}
 \left[M\frac{\Gamma(M+1+p)}{\Gamma(M+1)}
 +(M+1)\frac{\Gamma(M+p)}{\Gamma(M)}\right].
\end{equation}
We introduce the generating function of the expression in brackets:
\begin{equation}
 f(z)=\sum_{M\geq1}\left[M\frac{\Gamma(M+1+p)}{\Gamma(M+1)}
 +(M+1)\frac{\Gamma(M+p)}{\Gamma(M)}\right]z^M.
\end{equation}
Using the identity
\begin{equation}
 \sum_{M\geq1}\frac{\Gamma(M+p)}{\Gamma(M)}z^M
 =\Gamma(p+1)\frac{z}{(1-z)^{p+1}},
\end{equation}
and differentiating to obtain the factors $M$, we find
\begin{equation}
 f(z)=\Gamma(p+1)
 \frac{(p+3)z+(p-1)z^2}{(1-z)^{p+2}}.
\end{equation}
Its leading singular term at $z=1$ is
$2\Gamma(p+2)(1-z)^{-p-2}$. By \eqref{eq:leading-transfer},
\begin{equation}
 [z^M]f(z)=2M^{p+1}+\mathcal O(M^p).
\end{equation}
For even $N=2M$, the two candidate-cost numerators are
\begin{align*}
 &M\frac{\Gamma(M-1+p)}{\Gamma(M-1)}
 +(M-2)\frac{\Gamma(M+1+p)}{\Gamma(M+1)}
 +2\frac{\Gamma(M+p)}{\Gamma(M)},\\
 &(M-1)\frac{\Gamma(M-1+p)}{\Gamma(M-1)}
 +(M-1)\frac{\Gamma(M+1+p)}{\Gamma(M+1)}
 +2\frac{\Gamma(M+p)}{\Gamma(M)}.
\end{align*}
Their generating functions have the same leading singular term:
finite shifts of the index do not change it, and terms without a factor
$M$ have lower singular order. 

\subsection{\bf $-1<p<0$} 
The exact average cost is
\begin{equation}
 \mathbb{E}\left[E(\hat{h}_{\M})\right]
 =\frac{\Gamma(N+1)}{\Gamma(N+p+1)}
 \left[(N-1)\Gamma(p+1)+\frac{\Gamma(N+p-1)}{\Gamma(N-1)}\right].
\end{equation}
The generating function of the expression in brackets is
\begin{equation}
\begin{split}
 f(z)&=\sum_{N\geq2}\left[(N-1)\Gamma(p+1)
 +\frac{\Gamma(N+p-1)}{\Gamma(N-1)}\right]z^N\\
 &=\Gamma(p+1)z^2\left[(1-z)^{-2}+(1-z)^{-p-1}\right].
\end{split}
\end{equation}
Since $0<p+1<1$, the double pole dominates. 
Its coefficient is
$\Gamma(p+1)(N-1)$, while the second term has coefficients of order $N^p$.
The restriction $p>-1$ is necessary for the average to be finite;
for $p\leq-1$, the negative moments of adjacent spacings diverge.

\bibliographystyle{unsrtnat}
\bibliography{AssignmentANDTsp}
\end{document}